\documentclass{aa}  
\usepackage{graphicx}
\usepackage{txfonts}
\usepackage{amsmath}    % Advanced maths commands
\usepackage{amssymb}    % Extra maths symbols
\usepackage{natbib}
\usepackage{longtable}
\usepackage{float}
\usepackage{threeparttable}
\usepackage{subcaption}

\begin{document}

   \title{Recurrent low-level luminosity behaviour after a giant outburst in the Be/X-ray transient 4U~0115+63}

   \author{A. Rouco Escorial
          \inst{1,2}\fnmsep\thanks{alicia.rouco.escorial@northwestern.edu}
          \and
          R. Wijnands\inst{1}
          \and
          J. van den Eijnden\inst{1}
          \and
          A. Patruno\inst{3,4}
          \and
          N. Degenaar\inst{1}
          \and
          A. Parikh\inst{1}
          \and
          L.S. Ootes\inst{1}
          }

    \institute{Anton Pannekoek Institute for Astronomy, University of Amsterdam, Science Park 904, 1098 XH, Amsterdam, The Netherlands\\
    \and
    Center for Interdisciplinary Exploration and Research in Astrophysics (CIERA) and Department of Physics and Astronomy, Northwestern University, Evanston, IL 60208, USA\\
    \and
    Institute of Space Sciences (IEEC-CSIC) Campus UAB, Carrer de Can Magrans, s/n 08193 Barcelona, Spain\\
    \and
    ASTRON, the Netherlands Institute for Radio Astronomy, Postbus 2, 7900 AA, Dwingeloo, The Netherlands}
    
   \date{Received 2019; accepted 2020}

   \abstract{In 2017, the Be/X-ray transient 4U~0115+63 exhibited a new type II outburst that was two times fainter than its 2015 giant outburst (in the \textit{Swift}/BAT count rates). Despite this difference between the two bright events, the source displayed similar X-ray behaviour after these periods. Once the outbursts ceased, the source did not transit towards quiescence directly, but was detected about a factor of 10 above its known quiescent level. It eventually decayed back to quiescence over timescales of months. In this paper, we present the results of our \textit{Swift} monitoring campaign, and an \textit{XMM-Newton} observation of 4U~0115+63 during the decay of the 2017 type II outburst and its subsequent low-luminosity behaviour. We discuss the possible origin of the decaying source emission at this low-level luminosity, which has now been shown as a recurrent phenomenon, in the framework of the two proposed scenarios to explain this faint state: cooling from an accretion-heated neutron star crust or continuous low-level accretion. In addition, we compare the outcome of our study with the results we obtained from the 2015/2016 monitoring campaign on this source.}

\keywords{X-rays: binaries -- accretion -- stars: neutron --  pulsars: individual: 4U~0115+63}

\titlerunning{The faint luminosity behaviour of 4U~0115+63}
\authorrunning{Rouco Escorial et al.}
   \maketitle

\section{Introduction}\label{sec:P4_4U0115_2018_intro}

Be/X-ray transients (BeXBs) are X-ray binary systems composed of highly magnetised (with surface magnetic field strengths of $\sim$10$^{12-13}$\,G) neutron stars (NSs) that move around fast rotating Be stars in (often highly) eccentric orbits (see \citealt{Reig2011} for a review of such systems). The companion stars are B-type stars that show emission lines in their optical spectra (hence the additional `e' in their name) at some point in their lives. Eventually, these lines disappear, and regular B-type stellar spectra are observed from these stars (see the review by \citealt{Porter2003}). The origin of those lines is attributed to the presence of temporary gaseous equatorial discs (also called `decretion discs') formed by the expelled material from the rapidly rotating Be stars (see \citealt{Rivinius2013}).

The X-ray transient behaviour of BeXBs can be classified into two general types of phenomena (see \citealt{Stella1986}; \citealt{Okazaki2001}): type I and type II outbursts (the latter are also called `giant' outbursts). Type I outbursts, which typically occur periodically, last generally only a (small) fraction of the orbital period. They occur when the NS is at the periastron of the orbit and passes through the decretion disc. In this scenario, the NS can accrete matter from the disc and become bright in X-rays. The typical X-ray luminosities reached during these events are L$_\textnormal{X}$$\sim$10$^{36-37}$\,erg\,s$^{-1}$. In the case of type II outbursts, which can last considerably longer than an orbital period, the X-ray luminosities are usually an order of magnitude higher than those observed during the type I outbursts. X-ray luminosities for these events often reach L$_\textnormal{X}$$\geq$10$^{38}$\,erg\,s$^{-1}$. Occasionally, the X-ray luminosities observed during type II outbursts can even exceed the Eddington-limit luminosity for a NS (L$_\textnormal{X}$\,>\,2$\times10^{38}$\,erg~s$^{-1}$; e.g. \citealt{Tsygankov2017c, Tsygankov2018}). The physical mechanism that causes this phenomenon is still not well understood. However, some studies describe special combinations between the NS orbital plane and the structure of the Be-star decretion disc as a possible scenario behind these giant outbursts  (e.g. \citealt{Okazaki2002}; \citealt{Moritani2013}; \citealt{Martin2014a}; \citealt{Monageng2017}), while other studies investigate the possibility of Kozai-Lidov oscillations in the decretion disk as the main cause of this type of activity (e.g. \citealt{Martin2014b}, \citealt{Laplace2017}).

Owing to the high X-ray fluxes of the systems,  most BeXBs studies focus on the behaviour displayed by these systems at high X-ray luminosities (i.e., L$_\textnormal{X}$\,>\,10$^{36}$\,erg~s$^{-1}$), meaning when they are in outburst (e.g. see \citealt{Reig2011} and reference therein). Although several BeXB systems had been studied in the past while exhibiting lower luminosities (L$_\textnormal{X}$$\sim$10$^{34-35}$\,erg~s$^{-1}$; see \citealt{Motch1991}; \citealt{Rutledge2007}) or, even fainter, when they were in their so-called quiescent state (L$_\textnormal{X}$$\sim$10$^{32-33}$\,erg~s$^{-1}$; e.g. see \citealt{Negueruela2000}; \citealt{Campana2001b}; \citealt{Campana2002}; \citealt{Orlandini2004}). Only during the last few years, the low-luminosity behaviour of BeXBs has received much more attention resulting in numerous publications (e.g. \citealt{Rothschild2013}; \citealt{Doroshenko2014}; \citealt{Reig2014}; \citealt{Elshamouty2016}; \citealt{Wijnands2016}; \citealt{Rouco2017a}; \citealt{Rouco2018b}; \citealt{Tsygankov2017b}; \citealt{Tsygankov2019a}). It is clear from these publications that the behaviour of BeXBs at these low luminosities is quite diverse, and these studies highlight the importance of understanding the different physical scenarios behind the low-level luminosity stage where BeXBs usually spend most of their lifetime.

\subsection{The low-luminosity behaviour of BeXBs}

During the luminous episodes of BeXBs (L$_\textnormal{X}$$\geq$10$^{36}$\,erg~s$^{-1}$), matter is accreted onto the NS surface at high mass-accretion rates. In this scenario, matter penetrates the NS magnetosphere and is guided by the magnetic field towards the NS magnetic poles, where it is accreted onto the NS surface, thus creating hot spots (e.g. \citealt{Elsner1977}; \citealt{Ikhsanov2001}; \citealt{Lii2014}). Once the accretion rate decays and the outburst comes closer to its end, both the NS spin period (P$_\textnormal{spin}$) and magnetic-field strength become crucial components in the further evolution of the system (e.g. see the discussions in \citealt{Tsygankov2017a} and \citealt{Rouco2018b}). 

In the case of relatively fast spinning rotators (in general, P$_\textnormal{spin}$$\leq$10\,s), the accretion flow might get centrifugally inhibited due to the rotating NS magnetospheric barrier, and it is thought that material may be expelled from the system during this phase (called the `propeller effect'; e.g. \citealt{Illarionov1975}; \citealt{Romanova2004}; \citealt{DAngelo2010}). Such a mechanism starts operating at a luminosity (typically L$_{\textnormal{X}_\textnormal{prop}}$$\sim$10$^{35-36}$\,erg~s$^{-1}$, for relatively fast spinning NSs) that is determined by the magnetic field strength and P$_\textnormal{spin}$ of the NS (see discussion in \citealt{Campana2002}). In some cases, this mechanism might not be effective enough to expel matter, and material is still stored in a disc (in what is often called a `trapped' or `dead' disc) that remains coupled to the NS magnetic field even after the active accretion period has halted (e.g. \citealt{Syunyaev1977}; \citealt{Dangelo2012}; \citealt{Patruno2013}; \citealt{Dangelo2014}).

In both the propeller regime and the trapped disc scenarios, it is assumed that material does not reach the NS surface anymore. In this framework, very faint quiescent emission is expected, and indeed these systems are detected at luminosities of  only  L$_\textnormal{X}$$\sim$10$^{32-33}$\,erg~s$^{-1}$ (e.g. \citealt{Mereghetti1987}; \citealt{Roberts2001}; \citealt{Campana2002}; \citealt{Reig2014}). However,  the exact emission mechanism(s) behind this low-level X-ray emission is not entirely clear yet. In some sources, the detection of pulsations, short-term variability, and the observation of high-energy spectra show that matter is still reaching the NS surface. Therefore, in these systems, the propeller mechanism is not fully effective, leading to a leakage of material through the magnetic field lines onto the NS magnetic poles. Nevertheless, our current understanding of this process is very limited \citep[e.g.][]{Mukherjee2005,Orlandini2004,Doroshenko2014}. 

An alternative scenario to explain the detected radiation when accretion is thought to be centrifugally inhibited (thus no accretion of matter is supposed to happen onto the NS surface), is the cooling emission from an accretion-heated NS. In this scenario, during the outbursts, nuclear reactions are induced in the NS crust (e.g. \citealt{Haensel1990, Haensel2003, Haensel2008}; \citealt{Steiner2012}; \citealt{Lau2018}) as the crust is compressed due to the accretion of matter onto the surface. Due to the energy released by these reactions, the thermal equilibrium between the NS crust and core is disrupted, with the crust becoming significantly hotter than the core (e.g. \citealt{Rutledge2002}).  When the outburst is over, the crust cools down (with most of the heat flowing into the core, which will heat up, albeit marginally, during only one outburst) until the equilibrium between the crust and the core is restored once again. This crust cooling can be observed using sensitive X-ray instruments. When the crust cooling is over, thermal emission might still be detectable from the NS if its core is sufficiently hot (i.e. it has been heated significantly by accretion over many outburst cycles). The detection of this thermal surface emission (either from a cooling crust or directly related to the core temperature) allows us to study the properties of the crusts and cores, and, therefore, those of the physics of ultra-dense matter in NSs. Most of the studies have been performed for systems that harbour NSs with low magnetic fields ($B \sim 10^{8-9}$ G; see \citealt{Wijnands2017}, for a review). However, not much is known about how the presence of strong magnetic fields ($B\sim10^{12-13}$ G) may affect the thermal state of accreting NSs (both the crust as well as the core). So far, only a handful of systems have shown thermal emission from the surface which can likely be associated with the thermal state of the core \citep[e.g.][]{Campana2001a,Elshamouty2016,Tsygankov2017b}. In addition, only two BeXBs (4U~0115+63 and V~0332+53) have shown evidence of potential cooling emission from their NS crusts, however, possible contribution from low-level accretion of matter onto the NSs in these systems could not be ruled out (see \citealt{Wijnands2016}, and \citealt{Rouco2017a} for in-depth discussions).

\subsection{The BeXB 4U 0115+63}

The BeXB 4U~0115+63 harbours a highly magnetised NS (B$\sim$1.3$\times$10$^{12}$\,G; see \citealt{Raguzova2005}) with P$_\textnormal{spin}$$\sim$3.62\,s (\citealt{Cominsky1978}). The compact object orbits around its early-type companion (B0.2Ve; see \citealt{Negueruela2001}) every 24.3 days (\citealt{Rappaport1978}). The distance towards the source is typically assumed to be $\sim$7\,kpc (\citealt{Okazaki2001}). This distance was recently confirmed by the \textit{Gaia} mission (\citealt{Gaia2016, Gaia2018}), which establishes a best distance estimate of 7.2$^{+1.5}_{-1.1}$\,kpc (\citealt{Rouco2018b}). Although the first outburst observed from the source was in 1969 using the \textit{Vela-5B} satellite (\citealt{Whitlock1989}),  the object was officially discovered in 1972 using the \textit{UHURU} satellite  (\citealt{Giacconi1972}; \citealt{Forman1978}). Since then, the system has shown both periodic type I outbursts (although not at every periastron passage) and type II outbursts (see \citealt{Boldin2013}; \citealt{Nakajima2015b}; \citealt{Nakajima2016a}; \citealt{Nakajima2016b}; \citealt{Nakajima2017}). 

After its 2015 type II outburst, the source was found at an elevated luminosity \citep{Wijnands2016,Rouco2017a} compared to its known quiescent luminosity (\citealt{Tsygankov2017b}). Continued monitoring observations of the source showed that this elevated level was slowly decreasing over time. This behaviour, combined with the observed spectral evolution of the source, was interpreted as evidence for the cooling of the neutron-star crust that was heated due to the accretion of matter in the preceding outburst (see \citealt{Wijnands2016,Rouco2017a}; although emission due to a slowly decreasing accretion rate onto the NS could not be ruled out).

In July 2017, the source exhibited another type II outburst (\citealt{Nakajima2017}), which lasted approximately one and a half months. In our paper, we present the results of our monitoring campaign on 4U~0115+63 after this latest type II outburst using the Neil Gehrels \textit{Swift} observatory (\citealt{Gehrels2004}; cited as \textit{Swift} from now on), and, additionally, one \textit{XMM-Newton} observation. This type II outburst (and our subsequent monitoring of the source when the outburst was over) is an exceptional opportunity to further investigate the nature of the observed low-luminosity state in this source, and maybe to conclusively demonstrate (or disprove) that it is caused by the cooling of an accretion-heated crust.

\section{Observations, analysis, and results}\label{sec:P4_4U0115_2018_analysis}

The BeXB 4U~0115+63 showed a type II outburst in 2017 July-August. This outburst was monitored using both the Burst Alert Telescope (BAT; \citealt{Barthelmy2005}) and the X-ray Telescope (XRT; \citealt{Burrows2005}; see Table \ref{tab:P4_4U0115_2018_log_observations} for more details of the XRT observations used for our spectral analysis in Section \ref{subsubsec:P4_4U0115_2018_Swift_spectra}) on board \textit{Swift}. These instruments allowed us to monitor the giant outburst and its consequent decaying phase. The XRT also allowed us to study the subsequent transition to the low-luminosity state of the source (similar to our observational campaign on the source in 2015/2016; see \citealt{Wijnands2016}; \citealt{Rouco2017a}). In addition, we obtained an observation on 2018 January 23 using {\it XMM-Newton} (\citealt{Jansen2001}) to study this low-luminosity state in further detail. We also re-analysed two archival {\it XMM-Newton} observations of the source (see Table\,\ref{tab:P4_4U0115_2018_log_observations} for more information about the {\it XMM-Newton} data we used; see also \citealt{Tsygankov2017b} and \citealt{Rouco2017a} for the two archival observations).

\subsection{Light curve}\label{subsec:P4_4U0115_2018_lightcurve}

\begin{figure}[t]
	\center
	\includegraphics[width=\columnwidth]{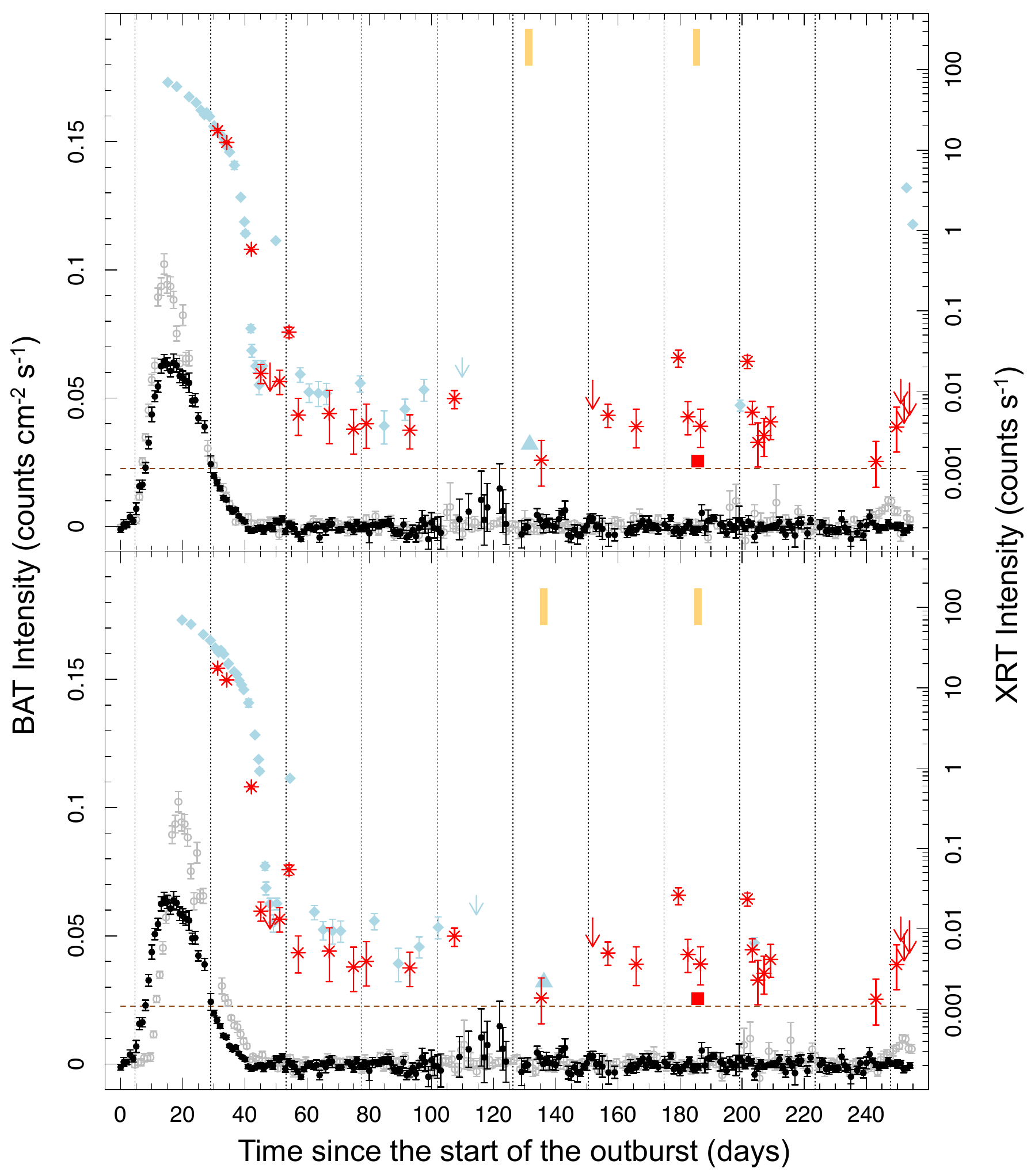}
	\caption{The \textit{Swift}/XRT (light blue diamonds) and \textit{Swift}/BAT (light grey open circles) light curves during and after the 2015/2016 type II outburst of 4U~0115+63 and the \textit{Swift}/XRT (red stars) and \textit{Swift}/BAT (black circles) light curves during the 2017/2018 campaign for the source. In the top panel, we show the light curves obtained during both campaigns but aligned using the start times of these outbursts. The zero-points of the light curves correspond to 2015 October 9 (MJD~57304), and 2017 July 22 (MJD~57956). In the bottom panel, we show the same light curves again but now aligned with respect to the occurrence of the first periastron passages after the type II outbursts (around day 50 in the light curves). Our 2016 and 2018 \textit{XMM-Newton} observations are shown as a light blue up-pointing triangle and a red square, respectively, and the times at which these observations were obtained are also indicated by the two yellow marks. The XRT and \textit{XMM-Newton} count rates are calculated in the $0.5-10$\,keV energy range, while the BAT count rates are shown in the $15-50$\,keV range. The \textit{XMM-Newton} count rates have been converted to \textit{Swift}/XRT count rates (see Sect.\,\ref{subsec:P4_4U0115_2018_lightcurve} for more details). Periastron passages for the 2017/2018 data set (computed using the orbital parameters in \citealt{Raichur2010}) are indicated by the vertical dotted lines, and the quiescent level of 4U~0115+63 (see Sec.\,\ref{subsec:P4_4U0115_2018_spectra}) is shown by the dashed brown line in the figures. All errors are 1$\sigma$.}
        \label{fig:P4_4U0115_2018_lightcurve}
\end{figure}

The XRT light curves  (see Figure\,\ref{fig:P4_4U0115_2018_lightcurve}) of the source were produced in the $0.5-10$\,keV energy range using the XRT web interface\footnote{\url{http://www.swift.ac.uk/user_objects/}} (\citealt{Evans2009}) with the source coordinates as given in \cite{Reig2015}. The BAT light curves  (also displayed in Figure\,\ref{fig:P4_4U0115_2018_lightcurve}) were obtained from the BAT transient monitor web page\footnote{\url{https://swift.gsfc.nasa.gov/results/transients/weak/4U0115p634/}} in the $15-50$\,keV energy range (\citealt{Krimm2013}). In addition, the \textit{XMM-Newton} count rates were converted to XRT count rates (also called `inferred XRT count rates') within the $0.5-10$\,keV energy range using the WebPIMMS tool\footnote{\url{https://heasarc.gsfc.nasa.gov/cgi-bin/Tools/w3pimms/w3pimms.pl}} and the spectral parameters of each \textit{XMM-Newton} observation (see Section \ref{subsubsec:P4_4U0115_2018_XMM-Newton_spectra} for more details about the {\it XMM-Newton} spectral analysis).

In Figure\,\ref{fig:P4_4U0115_2018_lightcurve}, we show the BAT and XRT light curves for both 2015/2016 and 2017/2018 observational campaigns, including the \textit{XMM-Newton} observations. Both panels in Figure\,\ref{fig:P4_4U0115_2018_lightcurve} display different representations of the same light curves but with different zero-points, depending on which date the outbursts started (top panel) or by matching the time of the periastron passages of both epochs (bottom panel; using the first mini type I outburst [see below] after the type II outburst to line up the two different data sets). The reason behind these different light-curve representations is to allow the reader to notice the recurrent low-luminosity behaviour of the source after the type II outbursts (Figure\,\ref{fig:P4_4U0115_2018_lightcurve}, top panel) and to see the detected variability close to periastron during both campaigns more clearly (Figure\,\ref{fig:P4_4U0115_2018_lightcurve}, bottom panel). The 2017 type II outburst lasted a similar number of days ($\sim$45) to the 2015 giant outburst (see BAT light curves in the top panel of Figure\,\ref{fig:P4_4U0115_2018_lightcurve}), but the BAT intensity of the former peaked at only about a half of the intensity observed during the 2015 outburst ($\sim$0.06\,counts~cm$^{-2}$~s$^{-1}$ versus $\sim$0.1\,counts~cm$^{-2}$~s$^{-1}$, respectively; see Figure\,\ref{fig:P4_4U0115_2018_lightcurve}). The XRT was used to monitor both outbursts, although the 2015 outburst was more densely covered. 

The 2015/2016 XRT campaign covered the decay of the outburst. During the final stages, the source showed a fast transition from $\sim$2.5\,counts~s$^{-1}$ to $\sim$1$\times$10$^{-2}$\,counts~s$^{-1}$ in four days, suggesting that the source possibly entered the propeller regime (see \citealt{Tsygankov2016} for more details). The 2017/2018 campaign consisted of fewer observations during the decay phase of the outburst, however, as one can notice in Figure\,\ref{fig:P4_4U0115_2018_lightcurve} (top panel), the XRT count rate during this outburst decreased from $\sim$0.6\,counts~s$^{-1}$ to $\sim$1.6$\times$10$^{-2}$\,counts~s$^{-1}$ in three days, meaning it shows a similar rapid decay as during the previous outburst, and suggesting that 4U 0115+63 also entered the proposed propeller regime during the 2017 outburst. The approximately similar count rates during both outbursts, at which this acceleration of the decay happened, suggest that it is a recurrent and likely fundamental property of the outburst behaviour of this source. 

A transition into the propeller regime would be consistent with the subsequent observations. In both cases, the source did not directly transit to quiescence, but it settled in a decaying low-luminosity state, approximately a factor of 10 above the quiescent level. In Figure\,\ref{fig:P4_4U0115_2018_lightcurve} (top panel), one can see that the low-luminosity state of the source is recurrent after the giant outbursts regardless of the differences in the source behaviour during the preceding outbursts (i.e. the peak brightness was different between the two outbursts). However, during the 2017/2018 low-luminosity state, the source appears to be slightly fainter than during this same state after the 2015/2016 outburst. 

Similar to what was observed during the 2015/2016 low-luminosity state (see \citealt{Wijnands2016} and \citealt{Rouco2017a}), the source exhibited  short-term (days) increases in count rate on top of the decay trend during the 2017/2018 low-luminosity state. These enhanced emission periods occurred typically close to periastron passages (see Figure\,\ref{fig:P4_4U0115_2018_lightcurve} bottom panel; the brightest occurrences have been called `mini type I' outbursts; \citealt{Wijnands2016}) and might be related to an increase of matter accreted onto the NS (\citealt{Campana2002}). Where exactly the matter is coming from, and how it is eventually accreted on to the NS remains unclear. Interestingly, similar to the 2015/2016 low-luminosity state, the source also exhibited such a mini type I outburst during the first periastron passage after the end of the 2017 outburst (Figure\,\ref{fig:P4_4U0115_2018_lightcurve} bottom panel; although this event was not as bright as the one observed after the 2015 outburst: $\sim$0.05 versus $\sim$0.7\,counts~s$^{-1}$, respectively). Our 2015/2016 XRT monitoring campaign stopped approximately 60 days after the end of the outburst, whereas in 2017 we monitored the source for longer. This allowed us to detect more of such mini type-I outbursts (Figure\,\ref{fig:P4_4U0115_2018_lightcurve} bottom panel). It is quite possible that after the 2015 outburst, the source also exhibited multiple of such events, but that these were missed because of the lack of XRT monitoring. Overall, the behaviour of the source after both type II outbursts is very similar, with respect to the general decay trend as well as the variability detected on top of this.

Our 2017/2018 monitoring campaign finished during a new periastron passage of the source at $\sim$250 days after the onset of the giant outburst. The source was barely detected during these observations (Figure\,\ref{fig:P4_4U0115_2018_lightcurve} bottom panel). We obtained a marginal detection of the source (three photons in 0.9\,ks; ObsID~00031172089) followed by three non-detections. When combining these three observations  (ObsIDs~00031172091, 00031172092, and 00031172093) the source was still not detected in a total exposure time of $\sim$2.48\,ks: resulting in a  2$\sigma$ upper limit of $<$3.5$\times$10$^{-3}$\,counts~s$^{-1}$ (obtained following the method described in \citealt{Gehrels1986}). The small number of photons detected from the source in ObsID~00031172089 did not allow us to perform any spectral fitting (following the criteria introduced in Sec.\,\ref{subsubsec:P4_4U0115_2018_Swift_spectra}).

%
%-----------------------------Figure Start------------------------------
%-----------------------------Figure End----------------------------
%

%
%-----------------------------Table Start------------------------------
\renewcommand{\arraystretch}{1.25}
\begin{table*}[t]
    \small
    \caption{Log of the \textit{Swift} and \textit{XMM-Newton} observations of 4U~0115+63 used for the spectral analysis in this paper.}
    \centering
    \begin{tabular}{lrcccc}
        \hline\hline
        \multicolumn{6}{c}{\textit{Swift}}  \\
        \hline
        ObsID   &   Date    &   MJD   &   Exposure Time   &   Count Rate  &   $\phi$  \\
            &       &    &   (ks)    &   (10$^{-3}$\,counts~s$^{-1}$) &   \\
        \hline
        \noalign{\smallskip}
        \multicolumn{6}{c}{\textit{2015/2016 campaign}}  \\
        000311720[30] & Nov 18  &   57344.1242$\pm$0.0055  &   $\sim$0.9   &   916$\pm$63               &   0.65    \\
        +[31]         & " 19    &   57345.889$\pm$0.037    &   $\sim$1.9   &   60.4$\pm$7.0             &   0.72    \\
        +[32]         & " 20    &   57346.18$\pm$0.14       &   $\sim$1.9   &   32.2$\pm$5.8             &   0.73    \\
        +[33]         & " 21    &   57347.487$\pm$0.038    &   $\sim$2.1   &   20.5$\pm$3.4             &   0.79    \\
        +[34]         & " 22    &   57348.549$\pm$0.038    &   $\sim$1.8   &   12.0$\pm$2.9             &   0.83    \\
        +[35]         & " 23    &   57349.681$\pm$0.039    &   $\sim$2.3   &   20.6$\pm$3.3             &   0.88    \\
        +[36]         & " 27    &   57353.894$\pm$0.071    &   $\sim$1.9   &   747$\pm$31               &   0.05    \\
        +[37]         & Dec 5   &   57361.761$\pm$0.069    &   $\sim$2.1   &   16.2$\pm$3.3             &   0.37    \\
        +[38]         & " 8     &   57364.589$\pm$0.038    &   $\sim$1.9   &   9.8$\pm$2.5              &   0.49    \\
        +[39]         & " 11    &   57367.6193$\pm$0.0063  &   $\sim$1.1   &   9.5$^{+3.7}_{-3.0}$      &   0.61    \\
        +[40]         & " 12    &   57370.21$\pm$0.14       &   $\sim$1.5   &   9.4$^{+3.1}_{-2.6}$      &   0.72    \\
        +[41]         & " 25    &   57381.067$\pm$0.032    &   $\sim$1.9   &   12.6$\pm$2.9             &   0.17    \\
        +[42]         & Jan 1   &   57388.782$\pm$0.036    &   $\sim$1.7   &   3.7$^{+2.0}_{-1.5}$      &   0.49    \\
        +[43]         & " 8     &   57395.49$\pm$0.43       &   $\sim$2.4   &   6.0$^{+2.0}_{-1.6}$      &   0.76    \\
        +[44]         & " 14    &   57401.60$\pm$0.37       &   $\sim$1.4   &   10.5$^{+3.6}_{-2.9}$     &   0.01    \\
        +[46]         & Apr 25  &   57503.37$\pm$0.31       &   $\sim$7.9   &   6.7$\pm$1.1              &   0.19    \\
        \noalign{\smallskip}
        \hline
        \multicolumn{6}{c}{\textit{2017/2018 campaign}}  \\
        000311720[63] & Sept 2  &   57998.055$\pm$0.036    &   $\sim$1.0   &   584$\pm$32              &   0.54    \\
        +[64]         & " 5     &   58001.1290$\pm$0.0056  &   $\sim$1.0   &   16.6$^{+5.0}_{-4.2}$    &   0.67    \\
        +[65]         & " 8     &   58004.0566$\pm$0.0058  &   $\sim$1.0   &   $<$13                   &   0.79    \\
        +[66]         & " 11    &   58007.1832$\pm$0.0049  &   $\sim$0.8   &   13.2$^{+5.1}_{-4.1}$    &   0.92    \\
        +[67]         & " 14    &   58010.1701$\pm$0.0055  &   $\sim$0.9   &   54.4$\pm$8.2            &   0.04    \\
        +[68]         & " 17    &   58013.121$\pm$0.029    &   $\sim$1.0   &   5.0$^{+3.1}_{-2.2}$     &   0.16    \\
        +[70]         & " 27    &   58023.1072$\pm$0.0031  &   $\sim$0.5   &   5.3$^{+5.0}_{-3.1}$     &   0.57    \\
        +[71]         & Oct 4   &   58030.9343$\pm$0.037   &   $\sim$1.1   &   3.4$^{+2.6}_{-1.7}$     &   0.89    \\
        +[72]         & " 9     &   58035.0880$\pm$0.0056  &   $\sim$0.9   &   3.9$^{+3.0}_{-2.0}$     &   0.06    \\
        +[73]         & " 23    &   58049.060$\pm$0.035    &   $\sim$1.8   &   3.3$^{+1.8}_{-1.4}$     &   0.64    \\
        +[74]         & Nov 6   &   58063.40$\pm$0.23       &   $\sim$2.7   &   8.1$\pm$2.1             &   0.23    \\
        +[75]         & Dec 4   &   58091.373$\pm$0.040    &   $\sim$2.8   &   1.38$^{+1.07}_{-0.72}$  &   0.38    \\
        +[77]         & " 25    &   58112.82$\pm$0.17       &   $\sim$2.4   &   5.0$^{+1.9}_{-1.5}$     &   0.26    \\
        +[78]         & Jan 3   &   58121.912$\pm$0.038    &   $\sim$1.2   &   3.6$^{+2.3}_{-1.7}$     &   0.63    \\
        +[79]         & " 17    &   58135.48$\pm$0.43       &   $\sim$0.9   &   26.2$\pm$6.2            &   0.19    \\
        +[80]         & " 20    &   58138.5689$\pm$0.0074  &   $\sim$1.3   &   4.8$^{+2.6}_{-1.9}$     &   0.32    \\
        +[81]         & " 24    &   58142.587$\pm$0.033    &   $\sim$1.2   &   3.7$^{+2.3}_{-1.7}$     &   0.48    \\
        +[83]         & Feb 8   &   58157.66$\pm$0.24       &   $\sim$1.9   &   23.5$\pm$4.4            &   0.10    \\
        +[84]         & " 10    &   58159.249$\pm$0.035    &   $\sim$2.1   &   5.5$^{+2.0}_{-1.6}$     &   0.17    \\
        +[85]         & " 12    &   58161.046$\pm$0.039    &   $\sim$2.3   &   2.3$^{+1.7}_{-1.2}$     &   0.24    \\
        +[86]         & " 14    &   58163.039$\pm$0.037    &   $\sim$2.0   &   2.8$^{+1.8}_{-1.3}$     &   0.32    \\
        +[87]         & " 16    &   58165.13$\pm$0.13       &   $\sim$1.8   &   4.2$^{+2.3}_{-1.7}$     &   0.41    \\
        \hline\hline
        \multicolumn{6}{c}{\textit{XMM-Newton}}  \\
        \hline
        0505280101    & 2007 Jul 21 &   54302.24$\pm$0.17  &   $\sim$7, $\sim$16, $\sim$16     &   13.0$\pm$1.5, 3.43$\pm$0.55, 3.75$\pm$0.56      &   0.45 \\
        0790180301    & 2016 Feb 17 &   57435.61$\pm$0.29  &   $\sim$15, $\sim$28, $\sim$29    &   24.8$\pm$1.3, 8.21$\pm$0.57, 10.10$\pm$0.62     &   0.41 \\
        0804940201    & 2018 Jan 23 &   58141.69$\pm$0.16  &   $\sim$6, $\sim$16, $\sim$15     &   15.3$\pm$1.8, 5.04$\pm$0.63, 5.13$\pm$0.66      &   0.45 \\
        \hline\hline
    \end{tabular}
    \tablefoot{The \textit{Swift}/XRT and {\it XMM-Newton} count rates are obtained for the $0.5-10$\,keV energy range and the errors are 1$\sigma$. The 2017/2018 XRT count-rate upper limit is calculated following the description in \citet{Gehrels1986}. The \textit{XMM-Newton} exposure times and count rates are shown in the following order: pn, MOS\,1 and MOS\,2. The effective exposure times (after filtering from background flaring events in the {\it XMM-Newton} data) are shown. In the last column, $\phi$ indicates the phase of the binary orbit at which the observations were obtained (periastron passage corresponds to $\phi$=0 and apastron to $\phi$=0.5). The errors on the MJD indicate the start and the end times of the interval during which the observations were obtained.}
    \label{tab:P4_4U0115_2018_log_observations}
\end{table*}
\renewcommand{\arraystretch}{1.0}
%-----------------------------Table End----------------------------
%

\subsection{Spectral analysis}\label{subsec:P4_4U0115_2018_spectra}

\subsubsection{\textit{Swift} observations}\label{subsubsec:P4_4U0115_2018_Swift_spectra}

For the {\it Swift}/XRT spectral analysis (see Table\,\ref{tab:P4_4U0115_2018_log_observations} for the observations used in this analysis), we used the 2017/2018 observations and, for consistency, we re-analysed the 2015/2016 data published by \citet{Rouco2017a}. As one can see in Figure\,\ref{fig:P4_4U0115_2018_lightcurve} (bottom panel), during the low-luminosity state, 4U~0115+63 exhibits variability which seems to be more prominent at periastron passage, suggesting that there might be enhanced accretion of matter onto the NS during periastron. Since this might alter the emitted spectra, we grouped our observations into two data sets depending on their orbital phases ($\phi$): the first set corresponds to the periastron-passage regime (if $0\leq \phi < 0.20$ and $0.80< \phi \leq 1.0$), and the second set to the out-of-periastron regime (if $0.20 \leq \phi \leq 0.80$). Due to the (very) low count rates in many of the observations obtained during the different out-of-periastron intervals (see Table\,\ref{tab:P4_4U0115_2018_log_observations}), we combined the data that were obtained during the same out-of-periastron interval\footnote{We calculated the dates of the combined observations as the weighted average (weighted by the exposure times) of the dates for each observation of the combined file. Errors are the start times and end times of the first and the last observations, respectively, in the combined files.} (see Table\,\ref{tab:P4_4U0115_2018_bbdy_spectral_results}) to guarantee a better signal-to-noise ratio.

We used the \textsc{heasoft} (v.6.17) software to analyse the XRT data, and we produced new cleaned XRT event files using the tool \textsc{xrtpipeline}. The source and background information (i.e. the count rates and spectra) were obtained using the tool \textsc{xselect}. As source extraction region, we used a circle of 15 pixels centred at the source position (\citealt{Reig2015}), while as background extraction region, we used an annulus with an inner radius of 60 pixels and an outer radius of 110 pixels. All the XRT observations were performed with the detector in Photon Counting (PC) mode. Only the count rates observed during the 2016 mini type I outburst (ObsID~00031172036) and during the 2017 decay (ObsID~00031172063) observations were above the pile-up limit ($\sim$0.5\,counts~s$^{-1}$), so we followed the standard thread\footnote{\url{http://www.swift.ac.uk/analysis/xrt/pileup.php}} to correct for this effect. We used \textsc{xrtpipeline} to create the exposure maps\footnote{For the combined data, we acquired their exposure maps following the method explained in \url{http://www.swift.ac.uk/analysis/xrt/exposuremaps.php}} and \textsc{xrtmkarf} to obtain the ancillary response files. The response matrix files (v.14) from the \textit{Swift} calibration database were used during this process and during the spectral fits.

Although our 2017/2018 observational campaign consisted of more observations than the previous monitoring campaign, the quality of the \textit{Swift} observations generally collected only a small number of counts. Therefore, we firstly opted for grouping the data into two counts per bin (with \textsc{grppha}) to avoid zero-values (or negative) in any of the spectral bins when subtracting the background spectra from the source ones. Additionally,  we only applied the spectral analysis to those observations that had four or more bins in their spectra. We used \textsc{xspec} (v.12.9.0)\footnote{\url{https://heasarc.gsfc.nasa.gov/xanadu/xspec/}} for our spectral analysis. We fitted the data in the $0.5-10$\,keV energy range using W-statistics (background subtracted C-statistics; \citealt{Wachter1979}). We used either an absorbed power-law model (\textsc{pegpwrlw}) or an absorbed black-body model (\textsc{bbodyrad}). This allows us not only to investigate the spectral evolution of our data, and, with that, the potential emission mechanisms, but also to compare our results with those in existing studies for this source (\citealt{Wijnands2016}; \citealt{Rouco2017a}). 

The quality of the spectra does not allow us to prefer one model over the other, and the data could also be well fitted using other single-component models (e.g. a neutron-star atmosphere model, as used in \citealt{Elshamouty2016} and \citealt{Rouco2018a}). However, we are (very) limited by the quality of our spectra, and therefore we prefer to continue our interpretation with the simple power-law and black-body models. The absorption column density (N$_\textnormal{H}$) was modelled using \textsc{tbabs} with \textsc{wilm} abundances (\citealt{Wilms2000}) and \textsc{vern} X-ray cross-sections (\citealt{Verner1996}). We fixed the N$_\textnormal{H}$ to 9$\times$10$^{21}$\,cm$^{-2}$ (\citealt{Kalberla2005}) as we could not constrain this parameter from our spectra (this N$_\textnormal{H}$ value was also used in previous works on this source, again allowing for a direct comparison; e.g. \citealt{Wijnands2016, Rouco2017a}). We adopted a distance to the source of 7.2\,kpc (\citealt{Rouco2018a}). In the case of the black-body fits,  the unabsorbed $0.5-10$\,keV fluxes were obtained using the convolution model \textsc{cflux}.

%
%-----------------------------Table Start------------------------------
\renewcommand{\arraystretch}{1.5}
\begin{table*}[t]
    \caption{Spectral analysis results obtained from the black-body model fitting for the out-of-periastron regime observations.}
    \centering
    \begin{tabular}{lcccc}
        \hline\hline
        ObsID   &   kT$_\textnormal{bb}$ & R$_\textnormal{bb}$ & F$_\textnormal{X}$ & L$_\textnormal{X}$\\
           &   (keV)   &   (km)  &   (10$^{-13}$\,erg~cm$^{-2}$~s$^{-1}$)  &   (10$^{33}$\,erg~s$^{-1}$)\\
        \hline
        \noalign{\smallskip}
        \multicolumn{5}{c}{\textit{2007}}  \\
        0505280101      &   0.384$_{-0.025}^{+0.026}$   &  0.421$_{-0.060}^{+0.072}$    &   0.745$_{-0.056}^{+0.058}$ &  0.462$_{-0.035}^{+0.036}$ \\
        \noalign{\smallskip}
        \hline
        \multicolumn{5}{c}{\textit{2015/2016}}  \\
        +[32-33]        &   0.693$_{-0.061}^{+0.072}$   &  0.506$_{-0.087}^{+0.103}$  &   12.0$_{-1.4}^{+1.5}$      &   7.46$_{-0.88}^{+0.96}$ \\
        +[37-40]        &   0.570$_{-0.053}^{+0.064}$   &  0.507$_{-0.095}^{+0.114}$   &   5.49$_{-0.69}^{+0.75}$    &   3.41$_{-0.43}^{+0.47}$ \\
        +[42-43]        &   0.42$_{-0.10}^{+0.20}$     &  0.69$_{-0.69}^{+0.67}$  &   2.85$_{-0.67}^{+0.79}$    &   1.77$_{-0.42}^{+0.49}$ \\
        0790180301      &   0.443$\pm$0.012   &  0.439$_{-0.026}^{+0.028}$    &   1.423$\pm$0.051           &  0.914$_{-0.031}^{+0.032}$ \\ 
        \noalign{\smallskip}
        \hline
        \multicolumn{5}{c}{\textit{2017/2018}}  \\
        +[64-65]         &   0.55$_{-0.11}^{+0.16}$       &  0.60$_{-0.23}^{+0.34}$     &   6.6$_{-1.5}^{+1.8}$        &   4.06$_{-0.93}^{+1.09}$  \\
        +[74]           &   0.46$_{-0.11}^{+0.20}$       &  0.61$_{-0.61}^{+0.48}$       &   3.55$_{-0.88}^{+1.05}$     &   2.20$_{-0.54}^{+0.65}$  \\
        +[75]           &   0.473$_{-0.060}^{+0.064}$     &  0.32$^{*}$             &   1.05$_{-0.45}^{+0.63}$     &   0.65$_{-0.28}^{+0.39}$  \\
        +[77-78]        &   0.70$_{-0.13}^{+0.18}$        &  0.222$_{-0.079}^{+0.119}$  &   2.49$_{-0.64}^{+0.76}$     &   1.54$_{-0.40}^{+0.47}$  \\
        +[80-81]        &   0.54$_{-0.18}^{+0.41}$        &  0.37$_{-0.37}^{+0.57}$       &   2.29$_{-0.80}^{+1.03}$     &   1.42$_{-0.50}^{+0.64}$  \\
        0804940201      &   0.461$_{-0.024}^{+0.026}$     &  0.316$_{-0.037}^{+0.042}$   &   0.900$_{-0.061}^{+0.064}$  &   0.559$_{-0.038}^{+0.040}$  \\
        +[85-87]        &   0.327$_{-0.067}^{+0.087}$     &   1.01$_{-1.01}^{+0.91}$    &   2.20$_{-0.57}^{+0.68}$     &   1.36$_{-0.35}^{+0.42}$  \\
        \noalign{\smallskip}
        \hline\hline
    \end{tabular}
    \tablefoot{The N$_\textnormal{H}$ was fixed to 9$\times$10$^{21}$\,cm$^{-2}$ (\citealt{Kalberla2005}). The unabsorbed X-ray fluxes (F$_\textnormal{X}$) and luminosities (L$_\textnormal{X}$) are calculated in the $0.5-10$\,keV energy range and using a distance of $\sim$7.2\,kpc (\citealt{Rouco2018a}). All errors are expressed for 1$\sigma$ confidence interval. The $^{*}$ means that this parameter  was fixed to the listed value during the fitting process.}
    \label{tab:P4_4U0115_2018_bbdy_spectral_results}
\end{table*}
\renewcommand{\arraystretch}{1.0}
%-----------------------------Table End----------------------------
%

%
%-----------------------------Table Start------------------------------
\renewcommand{\arraystretch}{1.5}
\begin{table*}
    \caption{Spectral analysis results for the power-law model fitting.}
    \centering
    \begin{tabular}{lcccc}
        \hline\hline
        ObsID   &   Regime    &   $\Gamma$    &   F$_\textnormal{X}$  &   L$_\textnormal{X}$    \\
            &       &       &   (10$^{-13}$\,erg~cm$^{-2}$~s$^{-1}$) &   (10$^{33}$\,erg~s$^{-1}$)    \\
        \hline
        \noalign{\smallskip}
        \multicolumn{5}{c}{\textit{2007}}  \\   
        0505280101      &     Out of periastron      &   2.73$\pm$0.19            &   1.28$_{-0.15}^{+0.17}$    &   0.792$_{-0.095}^{+0.105}$       \\   
        \noalign{\smallskip}
        \hline
        \multicolumn{5}{c}{\textit{2015/2016}}  \\
        000311720[30]$^{a}$    &     Out of periastron      &   0.435$\pm$0.089          &   938$_{57}^{+61}$          &   582$_{-36}^{+38}$              \\
        +[31]$^{a}$            &     "             &   1.52$\pm$0.25            &   45.6$_{-6.0}^{+6.9}$      &   28.3$_{-3.7}^{+4.3}$            \\
        +[32-33]               &     "             &   1.75$\pm$0.28            &   19.7$_{-2.5}^{+2.8}$      &   12.2$_{-1.5}^{+1.8}$            \\
        +[34-35]               &     Periastron    &   2.40$\pm$0.39            &   16.8$_{-2.3}^{+2.9}$      &   10.4$_{-1.4}^{+1.8}$            \\
        +[36]                  &     "             &   0.232$\pm$0.067          &   746$_{-34}^{+36}$         &   463$_{-21}^{+22}$               \\
        +[37-40]               &     Out of periastron      &   1.84$_{-0.34}^{+0.28}$   &   9.5$_{-1.2}^{+1.8}$       &   5.88$_{-0.77}^{+1.10}$          \\
        +[41]                  &     Periastron    &   2.36$_{-0.57}^{+0.61}$   &   11.0$_{-2.6}^{+4.2}$      &   6.8$_{-1.6}^{+2.6}$             \\
        +[42-43]               &     Out of periastron      &   2.3$\pm$1.2              &   5.8$_{-1.4}^{+4.7}$       &   3.57$_{-0.85}^{+2.92}$          \\
        +[44]                  &     Periastron    &   2.62$_{-0.76}^{+0.84}$   &   8.9$_{-2.8}^{+4.8}$       &   5.5$_{-1.7}^{+3.0}$             \\
        0790180301             &     Out of periastron      &   2.369$\pm$0.082          &   2.43$\pm$0.14             &   1.506$_{-0.084}^{+0.087}$       \\
        000311720[46]          &     Periastron    &   1.49$\pm$0.33            &   5.35$_{-0.89}^{+1.09}$    &   3.32$_{-0.55}^{+0.68}$           \\
        \noalign{\smallskip}
        \hline
        \multicolumn{5}{c}{\textit{2017/2018}}  \\
        000311720[63]$^{a}$   &     Out of periastron      &   0.496$\pm$0.093          &   585$_{-37}^{+40}$          &   363$_{-23}^{+25}$            \\
        +[64-65]              &     "             &   2.09$_{-0.73}^{+0.75}$   &   11.9$_{-2.7}^{+4.0}$       &   7.4$_{-1.7}^{+2.5}$         \\
        +[66]                 &     Periastron    &   4.02$_{-1.3}^{+1.6}$     &   64.0$_{-64}^{+301}$        &   40$_{-40}^{+187}$          \\
        +[67]                 &     "             &   1.87$_{-0.33}^{+0.34}$   &   41.6$_{-6.4}^{+7.2}$       &   25.8$_{-4.0}^{+4.4}$          \\
        +[74]                 &     Out of periastron      &   1.50$_{-0.97}^{+0.85}$   &   8.7$_{-3.2}^{+12.2}$       &   5.39$_{-2.0}^{+7.5}$          \\
        +[75]                 &     "             &   2.34$^{*}$               &   2.06$_{-0.89}^{+1.24}$     &   1.28$_{-0.55}^{+0.77}$          \\
        +[77-78]              &     "             &   2.08$_{-0.62}^{+0.67}$   &   4.0$_{-1.0}^{+1.4}$        &   2.46$_{-0.64}^{+0.85}$          \\
        +[79]                 &     Periastron    &   0.68$_{-0.61}^{+0.59}$   &   31.7$_{-9.3}^{+14.7}$      &   19.7$_{-5.7}^{+9.1}$          \\
        +[80-81]              &     Out of periastron      &   3.05$_{-1.6}^{+2.1}$     &   6.49$_{-3.3}^{+35.7}$      &   4.03$_{-2.1}^{+22.2}$          \\
        0804940201            &     "             &   2.34$_{-0.16}^{+0.18}$   &   1.28$_{-0.15}^{+0.17}$     &   0.93$_{-0.11}^{+0.12}$          \\
        +[83]                 &     Periastron    &   0.70$_{-0.52}^{+0.51}$   &   28.1$_{-5.9}^{+8.0}$       &   17.4$_{-3.6}^{+5.0}$          \\
        +[84]                 &     "             &   1.20$_{-0.90}^{+0.91}$   &   5.1$_{-1.7}^{+2.8}$       &   3.2$_{-1.1}^{+1.7}$          \\
        +[85-87]              &     Out of periastron      &   3.83$_{-0.83}^{+0.99}$   &   6.1$_{-2.7}^{+6.8}$       &   3.8$_{-1.7}^{+4.2}$          \\
        \noalign{\smallskip}
        \hline\hline
    \end{tabular}
    \tablefoot{The N$_\textnormal{H}$ was fixed to 9$\times$10$^{21}$\,cm$^{-2}$ (\citealt{Kalberla2005}). The F$_\textnormal{X}$ and L$_\textnormal{X}$ are calculated in the 0.5$-$10\,keV energy range and using a distance of $\sim$7.2\,kpc (\citealt{Rouco2018a}). All errors are expressed for 1$\sigma$ confidence interval. The $^{*}$ indicates that the parameters was fixed to the given value during the spectral fits. The observations indicated with the $^{a}$ were obtained during the decay phase of the giant outbursts and not during the low-luminosity states. The results for these observations are listed for comparison.}
    \label{tab:P4_4U0115_2018_pwl_spectral_results}
\end{table*}
\renewcommand{\arraystretch}{1.0}
%-----------------------------Table End----------------------------
%

\begin{figure}
        \includegraphics[width=\columnwidth]{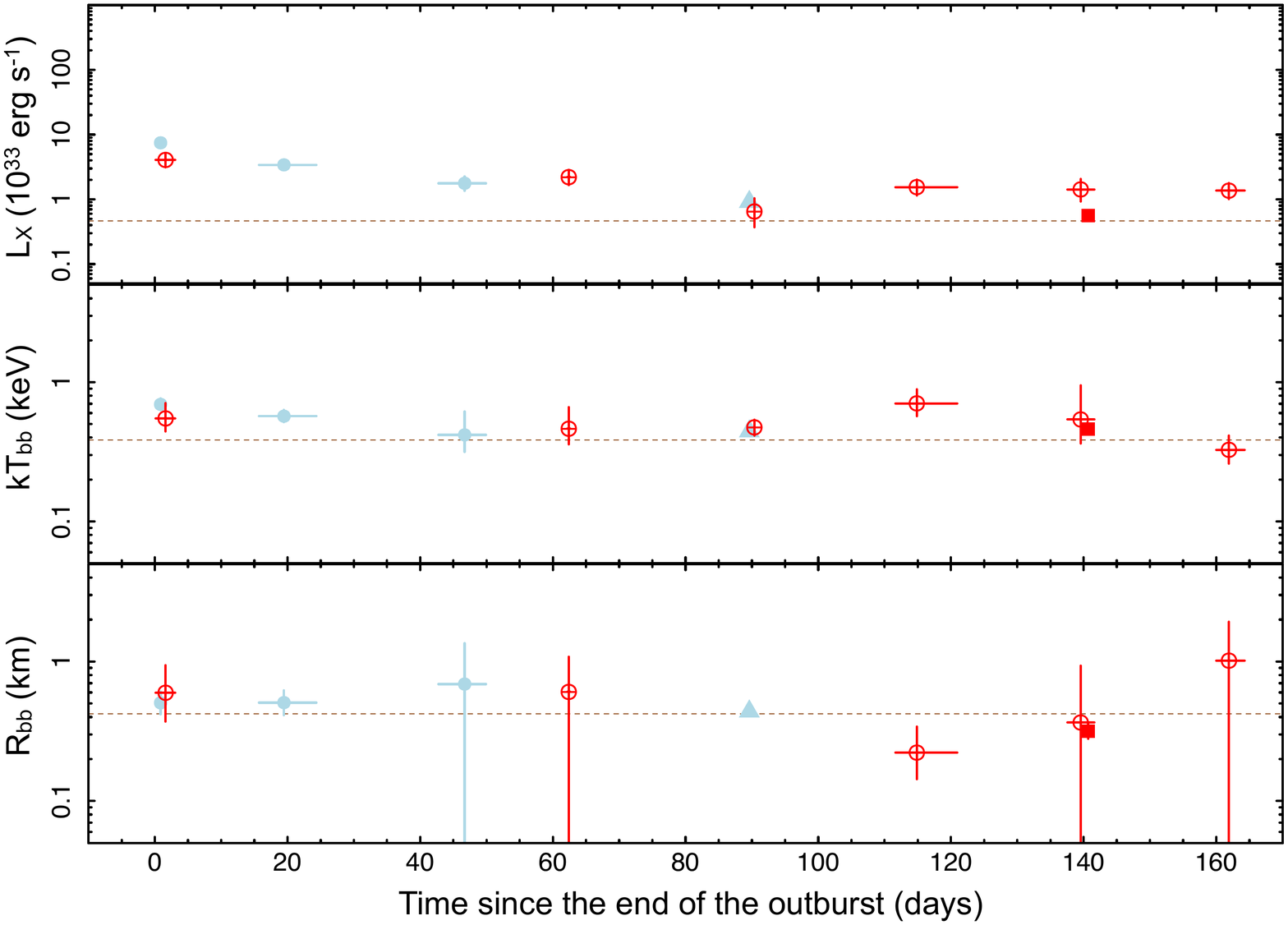}
    \caption{Evolution of black-body model spectral parameters for observations obtained during out-of-periastron regime. From top to bottom: the X-ray luminosity ($0.5-10$\,keV), the black-body temperature, and the corresponding emission region radii (values that were fixed during the spectral fits are not shown; see Sect.~\ref{subsubsec:P4_4U0115_2018_Swift_spectra}). The zero-points correspond to the dates of the end of the outbursts: 2015 November 20 (MJD~57346) for the 2015/2016 campaign and 2017 September 5 (MJD~58001) for the 2017/2018 campaign. The blue and red colours indicate the 2015/2016 and 2017/2018 data sets, respectively. Our \textit{XMM-Newton} observations are shown with an up-pointing triangle (light blue; 2015/2016 campaign) and a square (red; 2017/2018 campaign). Errors are 1$\sigma$ confidence levels. In some cases, the errors are smaller than the symbol. The horizontal dotted lines indicate the representative values measured during the 2007 quiescent {\it XMM-Newton} observation.}
        \label{fig:P4_4U0115_2018_bbody}
\end{figure}

\begin{figure}
        \includegraphics[width=\columnwidth]{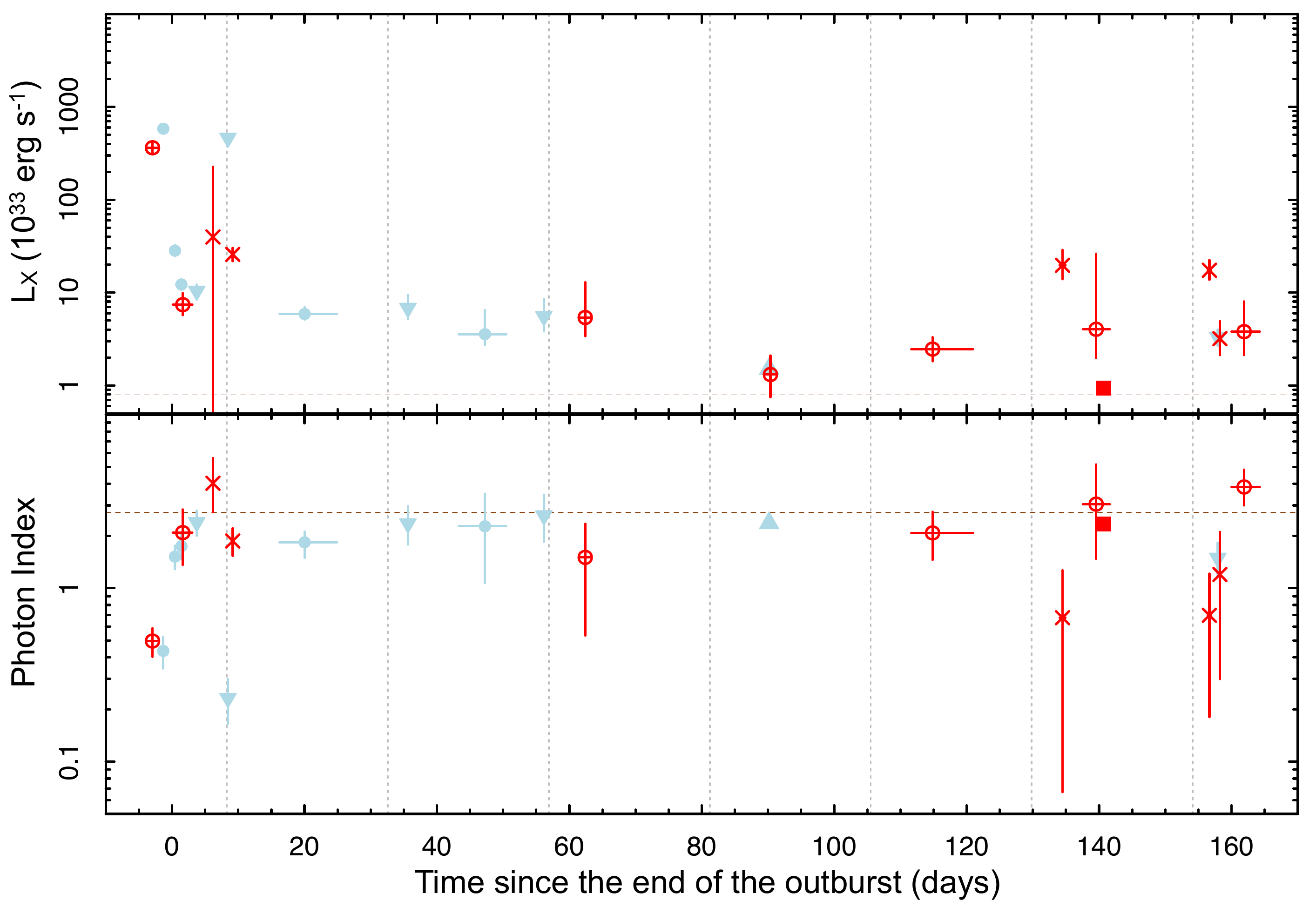}
    \caption{Evolution of spectral parameters for the power-law model. Top panel: time evolution of the X-ray luminosity ($0.5-10$\,keV energy range). Bottom panel: evolution of the photon index (values fixed during the spectral fits are not displayed; see Sect.~\ref{subsubsec:P4_4U0115_2018_Swift_spectra}). The zero-points are the same as those used in Figure\,\ref{fig:P4_4U0115_2018_bbody}, although the observations of the 2015/2016 campaign are shifted by a few days in order to line up the times of the periastron passages during this campaign with respect to the 2017/2018 campaign. The colours and symbols are the same as those used in Figure\,\ref{fig:P4_4U0115_2018_bbody}. Observations obtained in the periastron-passage regime are plotted with down-pointing triangles (light blue; in the case of 2015/2016 data) and crosses (red; for 2017/2018 observations). Extra observations during the decay of the giant outbursts are shown at negative days. Errors are 1$\sigma$ and may be smaller than the symbols used in the plot. Dotted vertical lines in the plots indicate the times of the periastron passages. The horizontal dotted lines indicate the representative values measured during the 2007 quiescent {\it XMM-Newton} observation.
    }
        \label{fig:P4_4U0115_2018_pwl}
\end{figure}

The results of our spectral fitting are shown in Table\,\ref{tab:P4_4U0115_2018_bbdy_spectral_results} and Table\,\ref{tab:P4_4U0115_2018_pwl_spectral_results}. Figures \ref{fig:P4_4U0115_2018_bbody} and \ref{fig:P4_4U0115_2018_pwl} show the time evolution of the spectral parameters for the black-body and power-law models, respectively. We apply the black-body model only to those observations obtained during the out-of-periastron regime, as the contribution due to possible low-level accretion is expected to be the lowest (and possibly negligible) during this orbital phase. In addition, the black-body model allows us to trace the evolution of the source temperature, which is a key parameter in the crust-cooling hypothesis. The black-body temperatures of the out-of-periastron intervals show a decay for both the 2015/2016  as well as for the 2017/2018 campaigns. The black-body temperature drops from  kT$_\textnormal{bb} \sim 0.69$\,keV to 0.44\,keV and from kT$_\textnormal{bb} \sim 0.55$\,keV to  0.33\,keV, respectively. The size of the emission region was between  R$_\textnormal{bb} \sim 0.3 - 1.0$\,km during both campaigns (see Table\,\ref{tab:P4_4U0115_2018_bbdy_spectral_results}; see also \citealt{Rouco2017a} for the spectral results obtained for the 2015/2016 campaign; our results are consistent with theirs). These sizes suggest that the energy is released from smaller regions than the NS radius, likely at the magnetic poles (which is consistent with the pulsations found during the {\it XMM-Newton} observations taken during these low-luminosity states; see Sect. \ref{subsec:P4_4U0115_2018_timing}). The source luminosity decayed in a similar way in both cases: from L$_\textnormal{X} \sim 7 \times 10^{33}$\,erg~s$^{-1}$ to $\sim 0.9 \times 10^{33}$\,erg~s$^{-1}$ in 2015/2016, and from L$_\textnormal{X} \sim 4 \times 10^{33}$\,erg~s$^{-1}$ to $\sim 0.6 \times 10^{33}$\,erg~s$^{-1}$ in 2017/2018 (see Figure\,\ref{fig:P4_4U0115_2018_bbody}). We note that the L$_\textnormal{X}$ inferred from the black-body fits to our 2017/2018 data are systematically lower than those obtained for the 2015/2016 data, suggesting that, although the general decay trend is seen during both campaigns, the source was systematically fainter in the 2017/2018 low-luminosity state than in 2015/2016. 

The results obtained from power-law fits (see Table \ref{tab:P4_4U0115_2018_pwl_spectral_results} and Figure\,\ref{fig:P4_4U0115_2018_pwl}) suggest that the spectra are harder during the 2015 mini type I outburst (ObsID~00031172036), the 2017 decay observation (ObsID~00031172063; probably due to an accretion tail from the main outburst), and the other two possible mini type I outbursts observed in 2018 (ObsID~00031172079 and ObsID~00031172083), than during the other intervals (with typical photon indices of $\sim 0.2-0.7$ versus $\sim 1.5-3$; see Table \ref{tab:P4_4U0115_2018_pwl_spectral_results}). Similar to our black-body fit results, the observed X-ray luminosities obtained from our power-law fits also show an overall decaying trend during both the 2015/2016 and the 2017/2018 campaigns, although, superimposed on this trend, we find enhanced emission close to the periastron passages (as can also be seen from the XRT light curves displayed in Figure \ref{fig:P4_4U0115_2018_lightcurve}; i.e. visible in the bottom panel). However, we do not find this enhanced X-ray emission at every periastron passage, so whatever the physical process behind these enhancements is, it is likely not active during each periastron passage.

%
%-----------------------------Figure Start------------------------------
%-----------------------------Figure End----------------------------
%

%
%-----------------------------Figure Start------------------------------
%-----------------------------Figure End----------------------------
%

\subsubsection{\textit{XMM-Newton} observations}\label{subsubsec:P4_4U0115_2018_XMM-Newton_spectra}
 
Using the {\it XMM-Newton} science analysis software or \textsc{SAS}\footnote{https://www.cosmos.esa.int/web/xmm-newton/sas}, we reduced and analysed our 2018 \textit{XMM-Newton} observation of 4U~0115+63 (ObsID~080494929), obtained $\sim$140 days after the start of the 2017 type II outburst, and re-analysed the other two previously obtained \textit{XMM-Newton} observations: the 2007 one obtained during quiescence (ObsID~0505280101; \citealt{Tsygankov2017b}) and the other low-luminosity state observation obtained in 2016 (ObsID~0790189391; \citealt{Rouco2017a}) after the 2015 giant outburst. We filtered the episodes of background flaring activity from the observations using the count rates in the $10-12$\,keV energy range for the EPIC-pn detector (from now on referred to as pn) and  $>$10\,keV for the EPIC-MOS detectors (hereafter MOS). We used the following count-rate thresholds (for ObsID~0505280101, ObsID~0790189391, and ObsID~080494929, respectively): for pn $\geq$0.4, $\geq$0.3, and $\geq$0.3\,counts\,s$^{-1}$; for MOS\,1 $\geq$0.25, $\geq$0.15, and $\geq$0.15\,counts\,s$^{-1}$; and for MOS\,2 $\geq$0.25, $\geq$0.2, and $\geq$0.2\,counts\,s$^{-1}$. We used the filtered files to produce calibrated event list files running the \textsc{emproc} and \textsc{epproc} tasks. We obtained the source counts and spectra using circular extraction regions with 20 arcsec radii on the source position, and we used source-free circular regions with 50 arcsec radii on the same CCD to extract the background counts and spectra. The source was faint during the {\it XMM-Newton} observations and our data were not affected by pile-up. We produced the response matrix and ancillary response files using the \textsc{rmfgen} and \textsc{arfgen} tasks, respectively. Finally, we grouped our data to two counts per bin using \textsc{specgroup} and performed the same spectral analysis as the one described in Sect.~\ref{subsubsec:P4_4U0115_2018_Swift_spectra} for the \textit{Swift} data. We tied the parameters between the pn, MOS\,1, and MOS\,2 spectra ,and we fitted the power-law and black-body models to them.

\begin{figure}
        \includegraphics[width=\columnwidth]{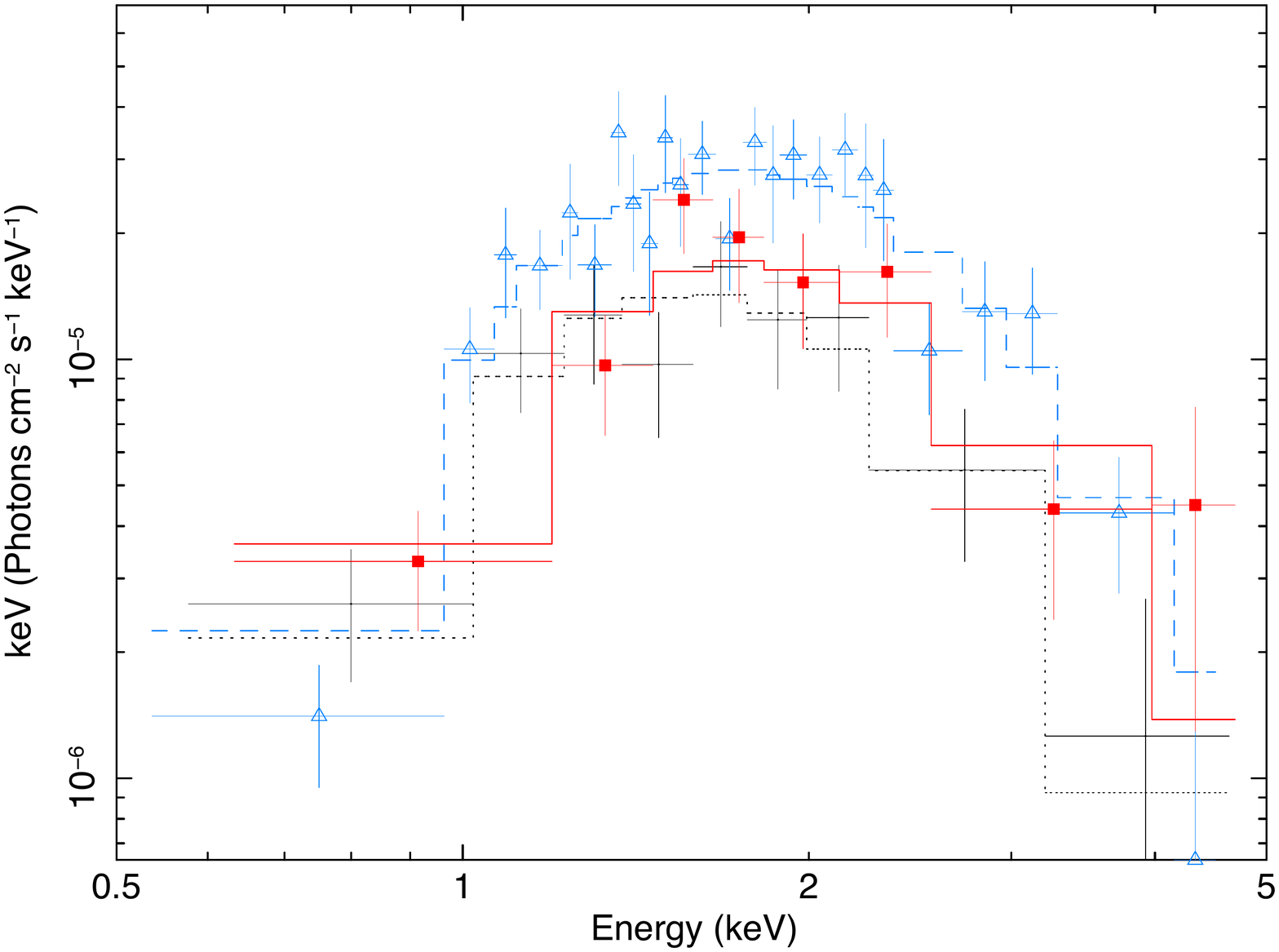}
    \caption{Comparison of unfolded pn spectra obtained during the three \textit{XMM-Newton} observations. All spectra were fitted using a black-body model. The spectra were rebinned for display purposes. The observations were obtained during quiescence in 2007 (black crosses and dotted line; ObsID 0505280101), $\sim$90 days after the end of the 2015 outburst (blue up-pointing triangles and dashed line; ObsID 0790189391), and $\sim$140 days after the end of the 2017 outburst (red squares and solid line; ObsID 0804940201).}
        \label{fig:P4_4U0115_2018_Unfolded_spectra}
\end{figure}

In Figure\,\ref{fig:P4_4U0115_2018_Unfolded_spectra}, we show the comparison of the three different epoch (2007, 2016, and 2018) pn spectra. As can be noticed, during the observation obtained in 2016 (blue point-up triangles) the source was clearly above its 2007 quiescent level (black crosses), while during the observation acquired in 2018 (red squares) the source was closer to quiescence. The results of our \textit{XMM-Newton} spectral analysis are displayed in Table\,\ref{tab:P4_4U0115_2018_bbdy_spectral_results} and Table\,\ref{tab:P4_4U0115_2018_pwl_spectral_results}, and the time evolution of the spectral parameters is shown in Figure\,\ref{fig:P4_4U0115_2018_bbody} for the black-body model, and in Figure\,\ref{fig:P4_4U0115_2018_pwl} for the power-law model. To determine if one of the models provided a better fit than the other, we followed the method used in \cite{Tsygankov2017b}. The difference between the C-values ($\Delta$C) obtained from the W-statistics of the power-law and black-body models indicates how preferable one model is over the other (i.e. when |$\Delta$C|>10; for more details see \citealt{Tsygankov2017b}). In the case of the three \textit{XMM-Newton} observations, |$\Delta$C| was higher than 10 for the power-law model compared to the black-body model, which demonstrates that the black-body model better describes the spectra. Nevertheless, we also report on the power-law spectral parameters for the \textit{XMM-Newton} observations in order to be able to compare the spectral shape during these observations with the shape observed during  the \textit{Swift} observations (as we could not statistically prefer one model over the other for these {\it Swift} spectra).

The black-body model provides best-fit spectral parameters with temperatures of kT$_\textnormal{bb}$\,$\sim$\,0.38$-$0.46\,keV and with small radii for the emission region, R$_\textnormal{bb}$\,$\sim$\,0.32$-$0.44\,km, (see Table\,\ref{tab:P4_4U0115_2018_bbdy_spectral_results}). If indeed this is the correct spectral model, this would suggest that we see thermal emission from the NS surface, likely from the magnetic poles. This is corroborated by the detection of pulsations from the source (see Sect.~\ref{subsec:P4_4U0115_2018_timing}). When fitting the spectra with a power-law model, the results show that the \textit{XMM-Newton} spectra are relatively soft, with photon indices of $\sim$2.3$-$2.7, also suggesting that the emission is thermal in nature. The observed spectra (using the {\it Swift}/XRT; Sect.~\ref{subsubsec:P4_4U0115_2018_Swift_spectra}) during the decay of the type II outbursts and during the mini type I outbursts were obtained when accretion of matter on to the NS surface was occurring at relatively high levels. These spectra are significantly harder than the low-luminosity state and quiescent spectra, with photon indices of $\sim$0.2$-$0.7 (see Table\,\ref{tab:P4_4U0115_2018_pwl_spectral_results}). During the {\it XMM-Newton} observations the source was detected at L$_\textnormal{X}$$\sim$0.9$-$1.5$\times$10$^{33}$\,erg~s$^{-1}$ ($0.5-10$\,keV; the lower and higher bounds are the black-body and power-law model X-ray luminosities, respectively) during the 2015/2016 low-luminosity state, while the luminosity was $\sim$0.6$-$0.9$\times$10$^{33}$\,erg~s$^{-1}$ in 2017/2018, which is almost the same as the observed quiescent level for the source, L$_\textnormal{X}$$\sim$0.5$-$0.8$\times$10$^{33}$\,erg~s$^{-1}$, in the 2007 observation. 

%
%-----------------------------Figure Start------------------------------
%-----------------------------Figure End----------------------------
%

\subsection{\textit{XMM-Newton} timing analysis}\label{subsec:P4_4U0115_2018_timing}

\begin{figure}
        \includegraphics[width=1.0\columnwidth]{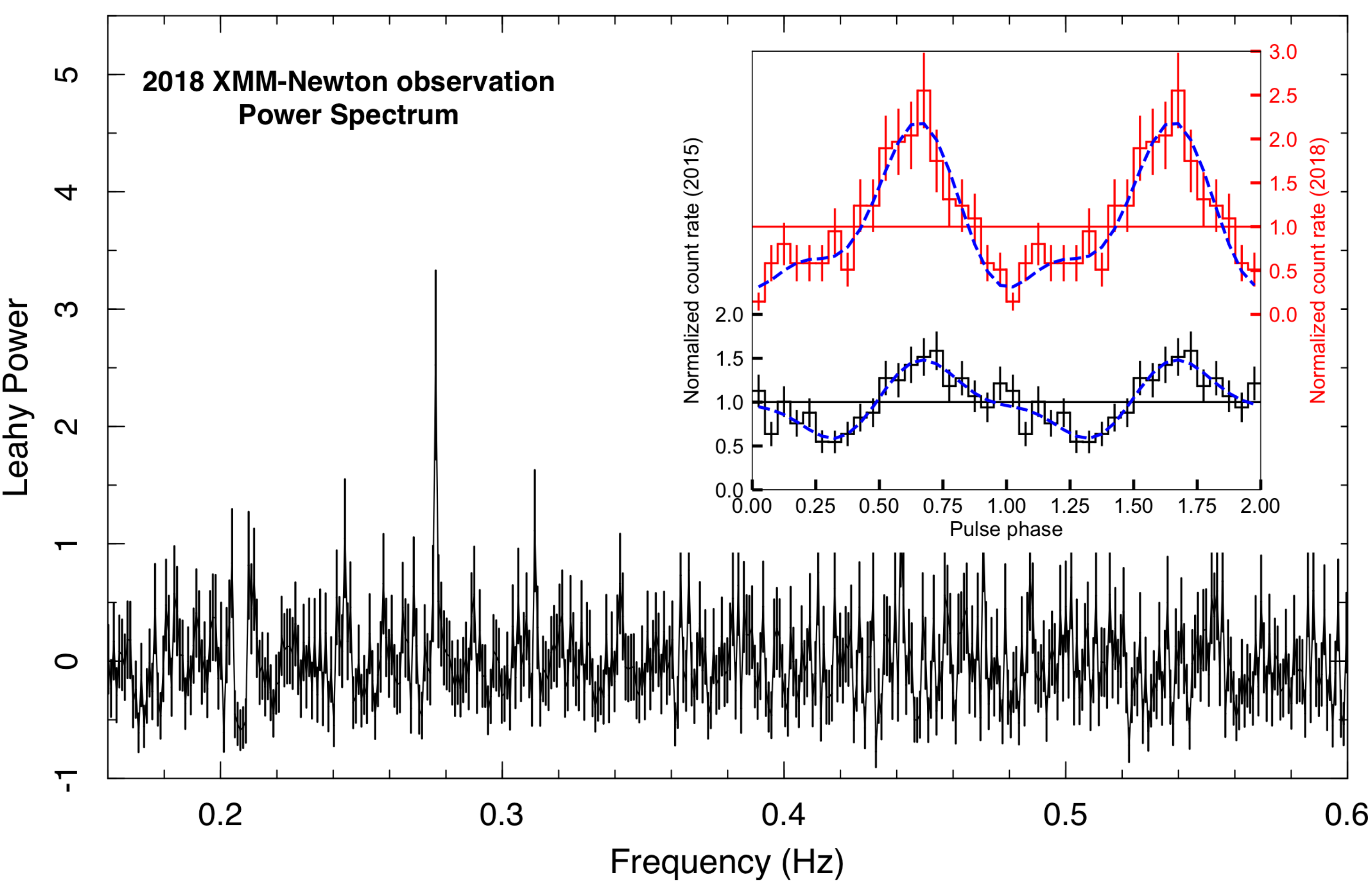}
    \caption{Final power spectrum (using the pn detector) created from our 2018 \textit{XMM-Newton} data (the final power spectrum from the 2015 \textit{XMM-Newton} observation was reported in \citep{Rouco2017a}). A clear pulsation is visible at $\sim$0.277\,Hz, which corresponds to the known NS spin frequency. Inset shows the pulse profiles obtained using the measured spin period during the 2015 (black, left axis) and 2018 (red, right axis) \textit{XMM-Newton} observations. The count rates are normalised using the mean values of the count rates. Two cycles of each pulse are displayed for clarity and both pulse profiles are aligned using the peaks of the profiles. The blue dashed lines show a sinusoidal model to fit the pulse profiles using two harmonics.}
        \label{fig:P4_4U0115_2018_Pulsations}
\end{figure}

For the timing analysis of our 2018 \textit{XMM-Newton} observation, we used the background filtered pn data and selected those X-ray photons that satisfied single and double patterns\footnote{\url{https://xmm-tools.cosmos.esa.int/external/xmm_user_support/documentation/uhb/epic_evgrades.html}} (i.e. patterns 0 to 4). We applied the same source extraction region as the one used during the spectral analysis (Sect.\,\ref{subsubsec:P4_4U0115_2018_Swift_spectra}). We 
applied the barycentric correction to the data using the \textsc{barycen} task. We rebinned our data to 0.1\,s time resolution and used 16384 points (1638.4 s of data) to produce the fast Fourier transforms (resulting in power spectra with a frequency range of 6$\times$10$^{-4}$$-$5\,Hz) using \textsc{ftools}. The resulting power spectra were normalised using the rms normalisation (power density units, (rms/mean)$^2$~Hz$^{-1}$) and averaged. The Poisson level was removed from the final power spectrum. A peak at the known NS spin frequency is clearly visible in the final power spectrum ($\sim$0.277\,Hz; see Figure\,\ref{fig:P4_4U0115_2018_Pulsations}).

%
%-----------------------------Figure Start------------------------------
%-----------------------------Figure End----------------------------
%

We folded the 2018 light curve (using a custom-made \textsc{python} script) around the spin period determined from our power spectrum (the measured spin is 3.61337$\pm$.00005\,s; the error corresponds to the standard deviation). To compare our timing results with the one we reported in \citep{Rouco2017a}, we applied the same folding technique to the 2015 \textit{XMM-Newton} light curve for consistency. The 2018 pulse profile varies over a range from 0.2 to 2.5, while the range of the 2015 pulse profile varies from 0.5 to 1.5 (both normalised to the mean values of the count rates; see inset in Figure\,\ref{fig:P4_4U0115_2018_Pulsations}). We used a sinusoidal function with two harmonics to fit the resulting pulse profiles (blue curves; Figure\,\ref{fig:P4_4U0115_2018_Pulsations}). We found that the fundamental amplitude is 79$\pm$7\,\% and that of the harmonic at twice of the fundamental frequency is 31$\pm$7\,\% for the 2018 pulse profile, while we obtained a fundamental amplitude of 37$\pm$5\,\% and harmonic amplitude of 14.1$\pm$4.6\,\% (1$\sigma$ errors; we note that these amplitudes are defined relative to the mean count rate) for the 2015 pulse profile. Therefore, we can consider these harmonics significant as the ratios between the pulse amplitudes and their errors are $>$3 (see \citealt{Patruno2010} for the reasoning behind using this criterion). The higher fundamental amplitude in 2018 than in 2015 implies that the difference in X-ray flux between the observation may be attributed to the non-pulsating component in the X-ray spectrum, however a detailed comparison and discussion of the pulse profiles are beyond the scope of our paper.

\section{Discussion}\label{sec:P4_4U0115_2018_discussion}

In this paper, we presented the results of our \textit{Swift}/XRT and \textit{XMM-Newton} observational campaign in 2017 and 2018 to monitor the evolution of the BeXB 4U~0115+63 after its type II outburst that occurred in July 2017. The aim of our study was to elucidate the mechanism(s) behind the low-luminosity state exhibited by the source after the type-II outburst in 2015, as the origin of the faint emission observed during this state has not been conclusively determined yet (\citealt{Campana2001a}; \citealt{Wijnands2016}; \citealt{Rouco2017a}). After the 2017 type II outburst, the source showed a behaviour very similar  to what was observed after the 2015 outburst. The system was again detected at a slowly decaying luminosity about a magnitude higher than what was observed in quiescence. This demonstrates that this low-luminosity state is a recurrent phenomenon in 4U~0115+63.

During both low-luminosity states of 4U~0115+63, there were episodes of (significantly) enhanced emission on top of the decaying low-luminosity trend. That is, after the end of both outbursts and at the time of the first periastron passage, the source exhibited the so-called mini type I outbursts. During the 2017/2018 low-luminosity state, several more of such events were found. However, it is very likely that such events were also present during the same state in 2015/2016, but we missed them since less intense monitoring was performed. These emission enhancements did not seem to alter the underlying decaying trend significantly. However, despite the many similarities, the main difference between both low-luminosity states is that the observed flux in the low-luminosity regime was slightly more elevated after the 2015 outburst than after the 2017 one. 

In the following two sections, we introduce the two most promising scenarios that may explain the origin of the observed behaviour of the source during its low-luminosity state: cooling emission from the surface of an accretion-heated NS (Sect.~\ref{subsec:cooling}), and low-level accretion of matter onto the neutron-star surface (Sect.~\ref{subsec:accretion}). One of the aims of our new monitoring campaign presented here, was to obtain more information to test which of the proposed scenarios could best explain the data. However, a combination of both scenarios seems to be the most likely explanation for the observed behaviour.

\subsection{Decaying cooling emission at a low-luminosity state} \label{subsec:cooling}

In the heating and cooling scenario, the NS crust in 4U 0115+63 has been heated (out of equilibrium with the core) due to accretion (i.e. during the type II outburst) and cools down (to re-establish equilibrium) once this accretion episode has halted. This scenario has been extensively discussed by \cite{Rouco2017a} in the context of the low-luminosity state after the 2015 type II outburst of the source, and we refer to that paper for details. Our \textit{Swift}/XRT monitoring campaign showed a very similar general decaying trend (only taking into account the data obtained during the out-of-periastron regime, and thus ignoring for now the mini type I outbursts and other potential emission contamination by accretion at periastron) in the X-ray luminosity after the 2017 outburst. This emission is soft, which indicates that it likely arises from the NS surface, and the detection of pulsations demonstrates that the emission is dominated by radiation originating from the hot spots on the surface (likely at the magnetic poles). 

The nearly identical behaviour seen after both type II outbursts suggests that the underlying physical process is something fundamental to this system. The hypothesis of the cooling of an accretion-heated NS crust is consistent with this inference and our new observations strengthen this scenario. However, despite the similar behaviour, the X-ray luminosities and temperatures during the 2017/2018 low-luminosity state are systematically slightly lower than the ones found in the 2015/2016 low-luminosity state. However, the 2017 outburst was a factor of two fainter than the 2015 outburst, so this difference in the behaviour during the low-luminosity state can easily be explained in the cooling scenario by postulating that, during the 2017 outburst, less heat was generated in the NS crust (since less matter was accreted) than during the 2015 outburst. Hence, the NS crust was heated to slightly lower temperatures during the 2017 outburst compared to the 2015 outburst.
 
When assuming that the low X-ray luminosity is due to the cooling emission from the NS crust, one can just fit a phenomenological model (an exponential decay function) to the cooling curve. This allows us to determine the characteristic cooling time, meaning to calculate the inferred time when the NS, heated by accretion, cools down. When fitting such an exponential decay function that levels off to a constant value to the temperature curve\footnote{Using the temperatures measured during the out-of-periastron regime XRT observations and during the {\it XMM-Newton} observations.} (the known quiescent temperature of the system), we obtained an e-folding time of 55$^{+13}_{-9}$ days for the low-luminosity state after the 2015 outburst. Unfortunately, it could not be constrained for the 2017 outburst, due to the large errors on the temperature values.  

We also calculated the cooling time for V0332+53, the other BeXB source for which the cooling of an accretion-heated NS crust might have been observed as reported in \citet{Wijnands2016}. Using the temperatures listed in that paper and the known quiescent temperature listed by \cite{Tsygankov2017b}, we obtained a cooling time of 48$^{+124}_{-33}$ days for V0332+53. This cooling time is similar to what we observed for the 2015/2016 campaign of 4U 0115+63, although the (positive) error bars are very large. Despite the large error bars, the NS crust appears to cool down faster in these two systems than the NS systems with low magnetic fields reported on by \citet{Homan2014}. Their Table 7 showed that the fastest observed cooling time scale for these kind of systems was $\sim$160 days, while others have characteristic cooling time scales of 200 to 500 days. This indicates that systems with high magnetic fields might (on average) cool down significantly faster than those with low magnetic fields, but we cannot draw strong conclusions with only two systems in our sample (also taking into account that the cooling scenario for systems with strong magnetic fields still needs to be confirmed)\footnote{We note that so far no crust cooling has been observed for the BeXB GS~0834-430 (due to lack of monitoring observations after its outbursts). However, a quiescent observation taken around $<$1\,year after the end of one of its outbursts, showed the source already back to full quiescence \citep{Tsygankov2017b}. This indicates that, also for this source, the crust-cooling timescale appears to be relatively short, similar to the other two sources discussed in this paper (in case that the crust was heated during its type II outburst, which might not necessarily be the case for all BeXBs; see, e.g. the lack of such a crust cooling in GRO~J1750-27 \citep{Rouco2018b}}. However, the tentative difference is intriguing and might point to a clear sign of the effect of a strong magnetic field on the heating and cooling behaviour of an accreting NS. A caveat in these studies is that we can only trace the behaviour of the hot spots, not of the full NS surface. This surface could have a lower temperature, which is very difficult to probe with our current data sets, however, the total emitted cooling radiation from the NS might be dominated by this surface emission (see discussions in \citealt{Elshamouty2016} and \citealt{Rouco2017a}). The cooling timescale of the rest of the NS surface might be significantly longer than from the hot spots. Indeed, this could bring the systems with high magnetic fields more in line with the ones with low magnetic fields.

\subsection{Contribution of low-level accretion} \label{subsec:accretion}

During the low-luminosity state of 4U 0115+63, the source is thought to be in the propeller regime, during which the accretion is centrifugally inhibited and material is expected not to reach the NS surface anymore. However, evidence against this scenario has been found in the low-luminosity state for some systems, such as detection of strong aperiodic variability and hard energy spectra (e.g. \citealt{Rothschild2013,Orlandini2004,Doroshenko2014}), or the detection of re-brightening events during the periastron-passage regimes (the mini type I outbursts) once the bright outburst finished (e.g. \citealt{Campana2001a,Campana2002,Wijnands2016,Rouco2018a}). One should not confuse this evidence with the detected pulsations and hard spectra found for the slowly spinning NS Be/X-ray transients (P$_{spin}\geq$100\,s) during their low-luminosity states, when the compact objects are accreting from non-ionised, cold accretion discs (e.g. \citealt{Tsygankov2017a,Tsygankov2019b,Rouco2018a}). In the case of fast rotators at low X-ray luminosities, the centrifugal barrier may not always be fully effective, and material can still be accreted, in small amounts, onto the NS surface. Therefore, it cannot be excluded that our target, 4U 0115+63, was also continuously accreting during all our observations in the low-luminosity state. Moreover, as already intensively discussed by \citet{Rouco2017a}, we cannot exclude conclusively that the general decay trend is not due to low-level accretion that slowly decreases in time. 

Our individual observations do not add extra information to what we had already learned from our 2015/2016 campaign, so they do not help to elucidate the physical mechanism(s) behind this slowly decaying low-luminosity state. However, the fact that we see a very similar decaying trend with similar decay time scales during both observational epochs, suggests that it is a fundamental property of the system (as also stated in Sect.~\ref{subsec:cooling}). It remains unclear if a slowly decaying accretion rate, with a very similar decay time scale, could explain the long-term luminosity trend. Furthermore, when we see conclusive evidence for accretion during the low-luminosity state of 4U 0115+63, it comes in the form of brief bursts of accretion with short-lived time scales (only a few days), which do not seem to affect the underlying cooling trend \citep[see][for further discussion about this]{Rouco2017a}

Strikingly, the accretion behaviour observed for 4U 0115+63 resembles that observed for V0332+53 \citep{Wijnands2016}, meaning the occurrence of the mini type I outbursts. However, 4U~0115+63 not only showed a mini type I outburst during the first periastron passage, but also after both the 2015 and the 2017 outbursts\footnote{Similarly, V0332+53 showed mini type I outbursts during the first two periastron passages after the end of its type II outbursts \citep{Wijnands2016}. However, after that the XRT monitoring stopped, so no information is available about the further behaviour of this source.}. During the 2017/2018 low-luminosity campaign, 4U~0115+63 also showed several more of these events (albeit less luminous) a few months after the end of the outburst. However, such accretion events were not observed at each periastron-passage regime for which we have data, demonstrating that whatever mechanism produces these mini type I outbursts, it is not always active at every periastron passage. Currently, it remains unclear how matter can overcome the centrifugal barrier during these accretion events (i.e. the accretion rate during these events is not high enough to overcome it) and reach the NS surface. In addition, it is not clear where this matter originally comes from. The most obvious source would be the decretion disc around the Be star. However, since only a very small amount of matter is accreted onto the NS, it remains to be determined if indeed such a small amount of matter can be transferred from the decretion disc of the star into the potential well of the NS.

\begin{acknowledgements}\label{sec:P4_4U0115_2018_acknowledgements}

ARE, AP, LSO and RW acknowledge support from an NWO Top grant, module 1, awarded to RW. JvdE and ND are supported by an NWO Vidi grant, awarded to ND. We are also grateful to the \textit{Swift} team for granting and scheduling our XRT observations. In addition, we acknowledge the use of public data from the \textit{Swift} data archive. This work is also based on observations obtained with \textit{XMM-Newton}, an ESA science mission with instruments and contributions directly funded by ESA Member States and NASA.

\end{acknowledgements}

%
%--------------------------------------------------------------------

\bibliographystyle{aa}
\bibliography{references}\label{references}

%
%--------------------------------------------------------------------

\end{document}